\documentstyle[aps,pra,epsfig,twocolumn]{revtex}
\begin{document}
\draft
\title{Stability of vortices in rotating traps: A 3D analysis}

\author{Juan Jos\'e Garc\'{\i}a-Ripoll, Victor Manuel P\'erez-Garc\'{\i}a}

\address{Departamento de M\'atem\'aticas, E.T.S.I. Industriales, \\ 
Universidad de Castilla--La Mancha,
13071 Ciudad Real, Spain.}

\date{\today}

\maketitle
\begin{abstract}
  We study the stability of vortex-lines in trapped dilute gases subject
  to rotation. We solve numerically both the Gross-Pitaevskii and the
  Bogoliubov equations for a 3d condensate in spherically and
  cilyndrically symmetric stationary traps, from small to very large
  nonlinearities.  In the stationary case it is found that the vortex
  states with unit and $m=2$ charge are energetically unstable. In the
  rotating trap it is found that this energetic instability may only be
  suppressed for the $m=1$ vortex-line, and that the multicharged vortices
  are never a local minimum of the energy functional, which implies that
  the absolute minimum of the energy is not an eigenstate of the $L_z$
  operator, when the angular speed is above a certain value, $\Omega >
  \Omega_2$.
\end{abstract}
\date{\today} 

\section{Introduction}

Since the first experimental realization of Bose-Einstein condensation
(BEC) in weakly interacting gases \cite{Anderson}, there has been a huge
theoretical and experimental effort to study its properties in the
framework of fully quantum theories and in the so called mean field limit
(Gross-Pitaevskii -GP- equations). These equations are formally Nonlinear
Schr\"odinger Equations (NLS) \cite{Perez-Garcia} which appear in many
fields of physics, e.g. in bulk superfluids and nonlinear optics to cite
only a few examples.

All of these physical systems have been long known to exhibit solutions
corresponding to topological defects \cite{Pitaevskii,Gross}, one of the
simplest being known as vortices (in two spatial dimensions) or
vortex-lines (in three spatial dimensions). Vortices are localized phase
singularities with integer topological charge which analogous in the
hydrodynamic interpretation to vortices those appearing in fluid dynamics
\cite{vortices}. In the framework of BEC studies it has been raised the
question of whether these nonuniform clouds of condensed gases may
support the existence of vortices in a stable form.

There is a huge literature on vortices and vortex properties in the
framework of NLS equations (including its particular cubic version, the
GP equation) and its nonconservative extensions, the Ginzburg-Landau (GL)
system, and vector GL models.  In particular the stability of $m$-charged
GP vortices in two dimensions was studied in \cite{Aranson}. In three
dimensions the GL case has been recently considered \cite{AransonII} and
vortex lines geometric instabilities have been found to strongly deform
the vortex lines. However the GP equation cannot be obtained as a limit
of the GL studied there since dissipation and diffusion are essential
ingredients of the models studied in Ref. \cite{AransonII}. This fact
makes the conservative case (GP) interesting by itself. Other analysis of
vortices and vortex stability in the framework of Nonlinear Optics are
included in Ref.  \cite{Kivshar}.

The current setups utilized to generate Bose-Einstein condensates use a
magnetic trap to confine the atomic cloud which is modeled by a parabolic
trapping potential. This is a distinctive feature from the common NLS
systems, in which the vortices are free and move in an homogeneous
background. The dynamics of a vortex in a spatially inhomogeneous two
dimensional GP problem was studied in Ref.  \cite{Rubinstein} using the
method of matched asymptotic expansions, but the authors did not consider
the stability of the 2D vortex itself. In principle, the vortex motion
equations \cite{Rubinstein} can be used to study the motion of a single
2D point vortex in spatially inhomogeneous GP problems.  However, the
dynamics of the many vortex case is more complicated and by no means
trivial. For simple approaches to the problem which do not include the
effect of vortex cores on the background field see Ref. \cite{Lund}. The
dynamics of 3D vortices is yet more complicated allowing the so-called
reconnection. To our knowledge there are no analytical results but only
qualitative numerical observations available \cite{Reconection}.  Another
theoretical framework where non-homogeneous dynamics of vortices has been
investigated is the possibility of pinning vortices in type-II
superconductors \cite{Chapmann}, but it is only the dynamics has been
considered through analytical approximation techniques with no comparison
with numerics. In all the previously discussed cases the vortex stability
is given for granted.

In the framework of Bose--Einstein condensed gases studies the problem of
the vortex stability has been considered in various papers that try to
solve the problem of linear and global stability, either from a purely
analytical point of view, such as in \cite{Stringari,Rokhsar,Fetter}, or
by mixing analytic and numerical techniques, \cite{Dalfovo,Dodd}.

In Ref. \cite{Dalfovo} the authors solve the Gross-Pitaevskii equation
and find the energies of the condensate in vortex states, for a number of
particles up to $N=10^4$. In Ref. \cite{Dodd} the authors solve the
Bogoliubov equations for an unit charge vortex in a stationary trap with
axial symmetry, their results being also limited to $N<10^4$. In Ref.
\cite{Rokhsar2} the authors address the problem of minimizing the energy
functional with a reduced basis of trial states that is only valid in the
limit of small $U$.

In this paper we unify and substantially extend what has been done in
previous works regarding these two questions: global energetic stability
and local stability of vortex states. First, in Sect.
\ref{sec:stationary} we solve the GPE for an axially symmetric harmonic
potential, with or without the action of a uniform magnetic field which
resembles the effect of a rotating trap. We calculate the lowest
stationary solutions that have a well defined value of the third
component of the angular momentum, $m=\langle L_{z}\rangle$, and we do
this for small and for very large values of the nonlinearity
($N\simeq10^7$). We find that there are continuous intervals of the
``angular velocity'', $[\Omega _{m},\Omega _{m+1})$, in which the
$m$-charged vortex state becomes energetically stable with respect to
other states of {\em well defined vorticity}. In Sect.
\ref{sec:linear-stability}, we study Bogoliubov's equations from two
different points of view: as a consequence of a linear stability analysis
of the Gross-Pitaevskii equation (GPE), and as the first corrections to
the mean field theory of the dilute condensate.  The concepts of
dynamical and energetic stability are defined, and it is demonstrated
that any possible destabilization of the system must be either of
energetic nature, or grow polynomially with respect to time. We next
solve the Bogoliubov equations for $m=1$ and $m=2$ unperturbed vortex
states in stationary traps. It is found that the $m=1$ and $m=2$ vortices
are only energetically unstable, which means that the lifetime of both
configurations is only limited by dissipation.  A similar treatment
reveals that rotation can only stabilize the unit charge vortex-line if
the angular speed is in a suitable range, $\Omega \in [\Omega_1,\Omega_2)$,
while outside of this range, $\Omega_2 < \Omega <\Omega_c$, the minimum of the energy functional is not an eigenstate of
the $L_z$ operator --i.e, it is {\em not symmetric under rotations}.
These results are confirmed by numerical simulations of the evolution of
perturbed vortices. In Sect.  \ref{sec:conclusions} we summarize our work
and discuss their implications.

\section{Vortex solutions of the GPE}
\label{sec:stationary}
\subsection{Stationary states of GPE in a uniform and constant magnetic
  field}

For small temperatures and small densities, the condensate is modeled by
the Gross-Pitaevskii equation (GPE) \cite{Pitaevskii,Gross}. We will
always refer to an axially symmetric trap with term that accounts for
rotation around the Z axis and may be generated by a weak magnetic field.
The form of the equation is
\begin{eqnarray}
\label{GPE-orig}
i\hbar \frac{\partial \psi }{\partial t} & = &
-\frac{\hbar ^{2}}{2m}\triangle \psi +\frac{1}{2}m\omega^2\left( \gamma ^{2}r^{2}+z^{2}\right) \psi \nonumber\\
& + & U_0N|\psi |^{2}\psi + \tilde{\Omega} L_{z}\psi .
\end{eqnarray}

Here $U_0 = 4 \pi \hbar^2 a/m$ characterizes the interaction and is
defined in terms of the ground state scattering length $a$. In all cases
we will take the normalization condition to be
\begin{equation}
  \label{normalization}
  \int |\psi |^{2}d^{3}x = 1.
\end{equation}

It is convenient to express Eq. (\ref{GPE-orig}) in a natural set of units
which for our problem is built up from two scales: the trap size (measured
by the width of the linear ground state), $a_{0}=\sqrt{\hbar/m\omega}$,
and its period, $\tau =1/\omega$.  With these definitions the equation
simplifies to
\begin{equation}
  \label{GPE}
  i\frac{\partial \psi }{\partial t}=\left[ -\frac{1}{2}\triangle
    +i\Omega \frac{\partial }{\partial \theta }+
    \frac{1}{2}(\gamma ^{2}r^{2}+z^{2})+U|\psi |^{2}\right] \psi ,
\end{equation}
while maintaining the normalization.

The new parameters, $\Omega = \hbar\tilde{\Omega}$ and $U=4\pi Na/a_{0}$,
represent the ``angular speed'' of the trap and the adimensionalized
interaction strength, respectively. For stability reasons (see below),
$\Omega$ will be of the order of magnitude of or smaller than the
strength of the trapping, $\omega$. The other parameter, $U$, will take
values from $0$ to $6\times 10^{4}$. As of the experiments with rubidium
and sodium, this implies a minimum of $10^{6} $ and a maximum of $10^{7}$
atoms which is in the range of current and projected experiments. The
shape of the trap is dictated by the geometry factor, and in this work it
will typically take two possible values: $\gamma =1$, corresponding to a
spherically symmetric trap, and $\gamma = 2$, corresponding to an axially
symmetric, elongated trap.

A stationary solution of (\ref{GPE}) will be of the form
$\psi(\vec{x},t)=e^{-i\mu t}\phi (\vec{x})$ \cite{Precision}, where $\mu$
may be interpreted both as a frequency and the chemical potential
\begin{equation}
  \label{stat-eq}
  \mu \phi =\left[ -\frac{1}{2}\triangle
    +i\Omega \frac{\partial }{\partial \theta }
    +\frac{1}{2}(\gamma ^{2}r^{2}+z^{2})+U|\phi |^{2}\right] \phi.
\end{equation}

Any solution of (\ref{GPE}) has an energy \emph{per} particle which is
given by the functional
\begin{eqnarray}
  \label{energy-functional}
  E(\psi ,N) & = &
  \int \left( \frac{1}{2}|\nabla \psi |^{2}
    -i\Omega \bar{\psi }\partial _{\theta }\psi \right) \nonumber \\
  & + & \int \left[ \frac{1}{2}\left( \gamma ^2r^2+z^2+U\left| \psi \right| ^{2}\right) \left| \psi \right| ^{2}\right] 
\end{eqnarray}
For a stationary solution it becomes
\begin{equation}
  \label{energy-per-particle}
  E(\psi,N) =\mu -\frac{U}{2}\int \left| \phi \right|^4.
\end{equation}
The stationary solutions of (\ref{GPE}) may also be interpreted as the
minimization of
\begin{equation}
  \label{constrained-E}
  {\cal L}_\mu = E(\psi,N) - \mu \int |\psi|^2
\end{equation}
subject to the constrain of Eq.  (\ref{normalization}). In that case
$\mu$ is nothing else but the Lagrange multiplier of the norm.

Since we are interested in single vortex solutions to the GPE we will
restrict our analysis to stationary states that are also eigenstates of
the $L_{z}$ operator. That is, we will look for solutions of the form
$\psi (r,z,\theta,t)=e^{-i\mu t}e^{im\theta}\phi (r,z).$ Summarizing, our
goal will be to find the unit norm functions $\phi^{(m)}_{\mu}(r,z)$ and
real numbers $\mu$ which are solutions of the equation
\begin{equation}
  \label{final}
  \mu \phi^{(m)}_{\mu} =\left[ -\frac{1}{2}\triangle - m \Omega
    +\frac{1}{2}(\gamma ^{2}r^{2}+z^{2})+U|\phi^{(m)}_{\mu} |^{2}\right] \phi^{(m)}_{\mu} ,
\end{equation}
Our treatment on the following sections will be fully three-dimensional
and no spurious conditions (e.g. periodicity) will be imposed on the
boundaries.  We want to obtain at least the lowest energy state for each
value of the vorticity, $m$. Also the dependency of spectrum with the
nonlinearity and the angular velocity, $\Omega$, is interesting since it
will allow us to find whether the vortex-line states may become
energetically favorable.

\subsection{Numerical method}

Due to the nonlinear nature of the problem we want to solve [Eq.
(\ref{final})] there are not many analytical tools available. The most
common (and maybe easiest) approach to the problem is to discretize the
spatial part and perform time evolution in imaginary time while trying to
preserve the normalization, a method which is related to the steepest
descent. The precision of the solution depends of the type of spatial
discretization used: finite differences (used for example in Refs.
\cite{Dalfovo,PerezII}) or spectral methods (such as the one used in Ref.
\cite{monopolo}). However these common methods, such as finite
differences \cite{Dalfovo} and similar spectral methods \cite{Dodd}, have
reached a maximum value of the interaction of $U=10^3$, which should be
contrasted with the value $U = 10^5$ that can be attained which the
technique to be presented later.

Properly speaking our technique is a Galerkin type method where one
performs the expansion of the unknown solution in a complete basis of
the Hilbert space under consideration. For convenience we have used
the basis of eigenstates of the harmonic oscillator with fixed
vorticity. In this basis our stationary solution is expressed as
\begin{equation}
  \psi^{(m)}_{\mu}(\vec{x},t)=e^{-i\mu t}e^{im\theta }\sum_{n}c_{n}P_n^{(m)}(r,z),
\end{equation}
Here the single index, $n$, denotes two quantum numbers, $(n_{z},n_{r})$,
regarding the axial and radial degrees of freedom, and $P_n^{(m)}$ is a
product of a Hermite polynomial, a Laguerre polynomial and a Gaussian
\begin{eqnarray}
  \mathletters
  P_n^{(m)} & = & C_n H_{n_z}(z) L_{n_r}(\rho^2) r^m e^{im\theta}
  e^{-(\rho^2+z^2)/2}, \\
  C & = & \sqrt{\frac{1}{\sqrt{\gamma}\sqrt{\pi} 2^{n_z} n_z!}}
  \sqrt{\frac{n_r!}{\pi \left(n_r+m\right)!}},
\end{eqnarray}
with $\rho=r/\sqrt{\gamma}$

Next, following the same convention about the indices, we have introduced
this expansion into Eq. (\ref{final}) to obtain
\begin{equation}
  \label{stat-discrete}
  \left( E^{ho}_{i}-\Omega m-\mu \right) c_{i}+U\sum _{jkl}A_{ijkl}^{(m)}\bar{c}_{j}c_{k}c_{l}=0,
\end{equation}
where $E^{ho}_{i}$ is the harmonic oscillator energy of the mode $P_{im}$
and the tensor $A^{ijkl}$ has the following definition
\begin{equation}
  \label{4-tensor}
  A_{ijkl}^{(m)}=2\pi \int \bar{P}_i^{(m)}\bar{P}_j^{(m)}P_k^{(m)}P_l^{(m)}drdz.
\end{equation}
Since the $P_i^{(m)}$ are products of known polynomials by exponentials
it could be possible, in principle, to evaluate the coefficients in the
tensor exactly with a Gaussian quadrature formula of the appropriate
order.  This approach was used in Ref. \cite{JIST} for the
three-dimensional case.  However, when one wishes to use a large number
of modes (which in our case is of about $1600$ for each value of $m$) to
achieve large nonlinearities, the search of the quadrature points becomes
more difficult than performing a stable integration by means of some
other methods, of which the simplest accurate one is Simpson's
rule\cite{Stoer}.

Once we fix all of the constants, $E^{ho}$, $A_{ijkl}^{(m)}$, $\mu$ and a
guess for the solution, it is feasible to solve (\ref{stat-discrete})
iteratively -e.g. by Newton's method\cite{Stoer}. However, it is wiser to
perform two simplifications before implementing the algorithm. The first
one is that all of the eigenfunctions, $P_n^{(m)} $, can be made real and
thus we can impose the coefficients in the expansion, $\{c_{n}\} $, to
also be real.

The second optimization is that, thanks to the symmetry of the problem,
the ground state of Eq.  (\ref{stat-eq}) has a well defined positive
parity. This allows us to eliminate redundant modes \cite{Note-1}, saving
memory and reaching higher energies and nonlinearities which otherwise
would be computationally hard to attain. On the other hand, we have
always checked that this method produced the same results as the complete
one for a selected and significant set of parameter values.

And finally it is important to note that the four--index tensor
(\ref{4-tensor}) is indeed a product of two smaller tensors,
corresponding to the integration on the $z$ and $r$ variables
\cite{JIST}. This decomposition is most important when working with a
large number of modes, because then the size of $A$ becomes extremely
large (i.e., $1600^4$ elements for $1600$ modes).

Concerning the evaluation of very high order polynomials as the ones
involved in our computations it is necessary to say that it is not a
simple task, specially for intermediate values of the spatial variables
since then there is a lot of comparable terms with usually different
signs and the cancellations induce numerical instabilities.  The usual
procedure to avoid this difficulty is to use Horner's method \cite{Stoer}
to evaluate the polynomial, which is comparable to using FFT techniques,
but in our case this is not enough and the evaluation of higher order
polynomials could be done only by the recursion formulas for the Hermite
and Laguerre polynomials.

We remark that the election of this spectral technique was largely
influenced by the need of reaching high nonlinearities which are not
achievable using the other approaches. Further details on the numerical
technique as well as convergence proofs will be given elsewhere
\cite{prepro}

\subsection{Results for the stationary states and spectrum}

\begin{figure}
  {\centering
    \resizebox*{0.9\columnwidth}{!}{\includegraphics{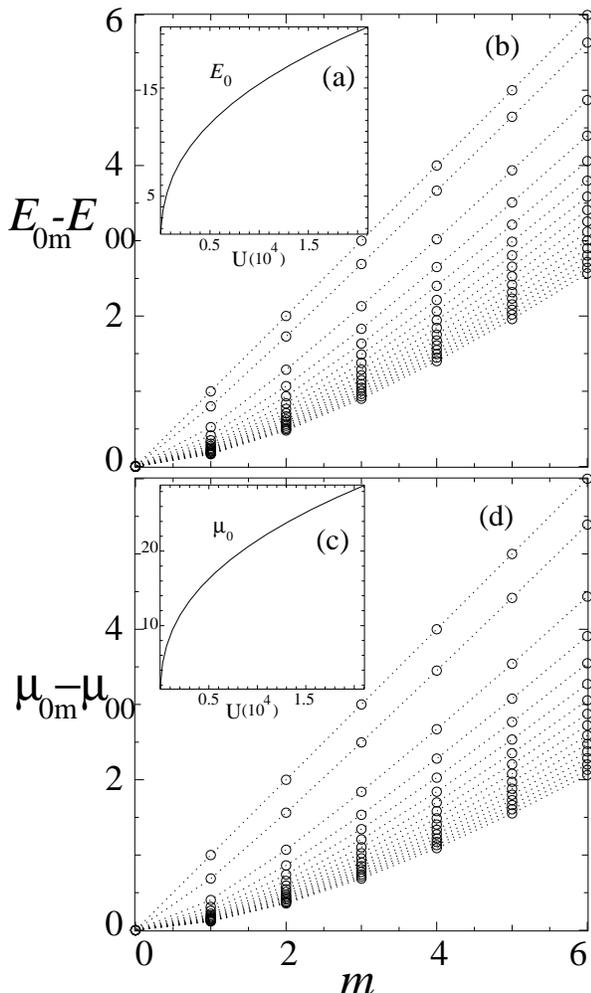}}
    \par}
  \caption{\label{spectrums-1} Plots (a) and (c) show the 
    ground state energy, \protect$E_{00}\protect $, and chemical
    potential, \protect$\mu _{00}(U)\protect $, dependence on the
    interaction strength.  Plots (b) and (d) show the chemical potential
    and the energy of the lowest state for each vorticity, always
    relative to value the ground state. The interaction values range
    from \protect$U=0\protect $ (upper diagonal) to \protect$
    U=50000\protect $ (lowest diagonal). All calculations shown
    correspond to the spherically symmetric trap, \protect$\gamma
    =1\protect $.}
\end{figure}

By using the preceding technique, we have searched the lowest states,
$(n_{z},n_{r}=0)$, for each branch of the spectrum with a different
vorticity, $m=0,\ldots, 6$.  This was performed for two geometries
corresponding to $\gamma =1$ (spherically symmetric trap) and $\gamma =
2$ (cigar shape trap), of a static trap, $\Omega =0 $, while varying the
intensity of the interaction from $0$ to approximately $50000$. The
results of this study are plotted in Fig.  \ref{spectrums-1}.

Remarkably, in the absence of rotation, and up from the lowest states,
both the spectrum and the energies can be fitted to a simple formula
\begin{equation}
  \label{fitted-energy}
  \mu _{0m}(N)=\mu _{00}(N)+\omega _{\text{eff}}(N)m.
\end{equation}
The first term is the chemical potential of the $m=0 $ ground state and is
not relevant to the dynamics. Using the Thomas-Fermi approximation one can
show that it grows proportionally to $\mu \propto N^{2/5} $, a behavior
which is approximately reflected in the numerical results shown in Fig.
\ref{spectrums-1}(c).

The second term is much more relevant to the evolution of the condensate.
It grows linearly, as the energy levels of a linear harmonic oscillator,
with an effective frequency, $\omega_{eff}(N)$, that decreases with the
interaction. The fact that the higher levels of the spectrum of $\mu$
remain equispaced even for large interactions is the reason why the
condensate exhibits an exponentially divergent response to the parametric
perturbation of the trap frequencies, as it is shown in Ref.
\cite{barrier-reson} and \cite{extended-reson}.

Now we want to study the stationary solutions in the presence of
rotation. For $\Omega \neq 0$ the proper functions with definite
vorticity remain the same, while the chemical potential and the energy
suffer a shift that depends on the vorticity of the state
\begin{equation}
  E_{nm}(U,\Omega )=E_{nm}(U,0)-m \Omega.
\end{equation}

\begin{figure}
  {\centering
    \resizebox*{0.9\columnwidth}{!}{\includegraphics{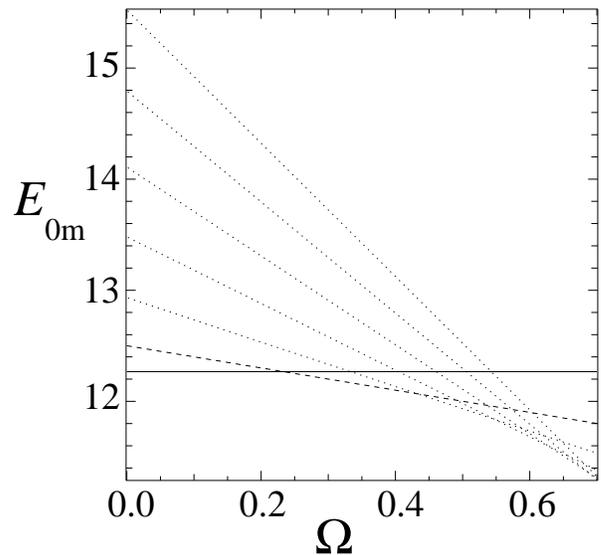}}
    \par}
  \caption{
    \label{rotation-1}Dependence of the energy levels on $\Omega$,
    \protect$E_{0m}(U,\Omega)\protect $, for a fixed value of the
    interaction strength, \protect$U\simeq 8000\protect$, and a
    spherically symmetric trap, \protect$\gamma =1\protect$. The
    horizontal line represents the vortex-free state, \protect$m=0\protect
    $, the dashed line the \protect$m=1\protect $ vortex state, and the
    dotted lines other multicharged vortex states.}
\end{figure}
This shift gives rise to an ample phenomenology which is pictured in Fig.
\ref{rotation-1}.First, we see that the degeneracy with respect to $m$ is
broken. The only other possible degeneracy that remains is with respect
to the $r$ and $z$ variables, but this will be removed for the case
without spherical symmetry, $\gamma \neq 1$.

Second, the $m=1,2,3...  $ branches of the spectrum become a minimum of
the energy functional with respect to other branches for continuous
intervals of the angular velocity, $[\Omega_m,\Omega_{m+1}]$, where
\begin{equation}
  \Omega_m = E_{0,m+1} - E_{0m}.
\end{equation}
However this does not mean that in these intervals the $m$-th vortex
state becomes a global minimum. Indeed, in Sect.
\ref{sec:linear-stability} we will only be able to prove that only the
$m=1,2$ vortex lines achieve the status of local minimum. It still
remains an open question under which situations the ground state must
have a well defined vorticity.

Third, even though the separation between the $m=0$ and $m=1$ states
becomes very narrow for large interactions, the stabilization frequency
$\Omega_{1}$ only approaches zero asymptotically with $U$. As a
consequence, $m=1$ states are never a global minimum of the energy in a
stationary trap, a fact that can be checked by just inspecting the energy
functional.

\begin{figure}
  {\centering
    \resizebox*{0.9\columnwidth}{!}{\includegraphics{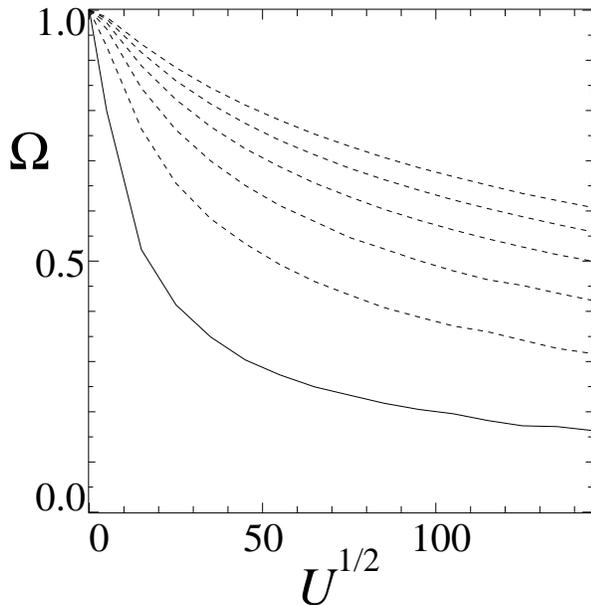}}
    \par}
  \caption{\label{omega-crit}
    Frequency of stabilization for the vortex states in a spherically
    symmetric trap, \protect$\gamma =1\protect $, as a function of the
    nonlinearity.  The lines are arranged in order of increasing
    vorticity, from \protect$ m=1\protect $ (solid line) to
    \protect$m=6\protect $ (dashed line on the top).}
\end{figure}

And finally, there is a critical value of $\Omega$ for which the energy
functional becomes unbounded by below [See Fig. \ref{rotation-1}] and
which coincides with the separation between energy levels for large
values of the vorticity. This critical value of the frequency,
$\Omega_c$, is such that all of the ground states for each value of the
vorticity have the same energy
\begin{equation}
  \label{collapse}
  E_{0m}(U,\Omega_c)=E_{0k}(U,\Omega_c),\ \forall k,m.
\end{equation}
Using Eqs. (\ref{collapse}) and a fit similar to the one in Eq.
(\ref{fitted-energy}), one finds that it is always smaller than the
critical frequency of the linear case
\begin{equation}
  \Omega _{c}=\omega _{\text{eff}}(U).
\end{equation}

\section{Stability of stationary states}
\label{sec:linear-stability}

\subsection{The linear stability equations}

In the preceding section we obtained stationary solutions of the mean
field model for the Bose-Einstein condensate, all of which had a well
defined value of the third component of the angular momentum operator. We
named those states vortices. In this section we now want to study the
stability of these solutions according to several criteria of a local
nature: local energetic stability and linear stability.

We begin our study from the adimensionalized Gross-Pitaevskii equation
(\ref{GPE}). First we expand the condensate wavefunction around a
stationary solution with a fixed vorticity.
\begin{eqnarray}
  \Psi (r,z,\theta ,t) & = &\psi_0 + \epsilon \psi_1 \nonumber \\
     & = & \left[ f(r,z)e^{im\theta }+
           \epsilon \alpha (r,z,\theta ,t)\right] e^{-i\mu(\Omega) t}.
\end{eqnarray}
We insert this expansion in Eq. (\ref{GPE}) and truncate the equations
up to ${\cal O}(\epsilon ^{1})$ thus getting
\begin{mathletters}
  \begin{eqnarray}
    i\partial _{t}\alpha & = &
    \left[ H_0+i\Omega\partial_\theta+2 U f^2\right] \alpha +
    U f^2 e^{-2im\theta }\bar{\alpha },\\
    -i\partial _{t}\bar{\alpha } & = &
    \left[ H_0-i\Omega\partial_\theta+2 U f^2\right] \bar{\alpha} +
    U f^2 e^{2im\theta }\alpha,
  \end{eqnarray}
\end{mathletters}
with $H_0=-\frac{1}{2}\triangle+\frac{1}{2}(\gamma^2 r^2 + z^2)
-\mu(\Omega)$. We can also write this equation in a more compact form
\begin{equation}
  \label{perturbed}
  i\frac{\partial}{\partial t}\vec{W} = \sigma_z {\cal H}(\Omega)\vec{W}
  = {\cal B}(\Omega) \vec{W}.
\end{equation}
by using the definitions
\begin{mathletters}
\label{definitions-1}
\begin{eqnarray}
  \vec{W} & = & \left(\begin{array}{c} \alpha \\ \bar{\alpha}\end{array}
                \right), \\
  \sigma_z & = & \left(\begin{array}{cc} 1 & 0 \\ 0 & -1 \end{array}
             \right), \\
  {\cal H}(\Omega) & = & H_0 + \left( \begin{array}{cc}
    i\Omega\partial_\theta + 2 U f^2 & U f^2 e^{-2im\theta} \\
    U f^2 e^{2im\theta} & - i\Omega\partial_\theta + 2 U f^2
  \end{array} \right).
\end{eqnarray}
\end{mathletters}

In the rest of this work we wish to study the dynamics that is involved
in Eq. (\ref{perturbed}). The simplest way to achieve this is to find a
suitable basis in which Eq. (\ref{perturbed}) becomes diagonal or almost
diagonal. In other words, we want a set of vectors,
$\vec{W}_k^t=(u_k(r),v_k(r))$ such that
\begin{equation}
  \label{diagonalization}
  \lambda_k \vec{W}_k = {\cal B} \vec{W}_k.
\end{equation}
If ${\cal B}$ has such a diagonal Jordan form, then the perturbation
evolves simply as
\begin{eqnarray}
  \vec{W} & =  &\sum c_k e^{i\lambda t}\vec{W}_k, \\
  \alpha(\vec{r},t) & = & \sum c_k e^{i\lambda t} u_k(\vec{r},t).
\end{eqnarray}
On the other hand, the lack of a diagonal form, or the existence of
complex eigenvalues leads to instability in a way that we will precise
later.

Associated to Eq. (\ref{perturbed}) there is an energy functional,
\begin{equation}
  \label{energy-perturbed}
  E_2(\alpha)=\int 2\bar{\alpha} H_0 \alpha + \psi_0^2\bar{\alpha}^2
              + \bar{\psi_0}^2\alpha^2 + 4|\psi_0|^2\alpha\bar{\alpha},
\end{equation}
and a constrained energy functional
\begin{equation}
  \label{constrained-E-perturbed}
  {\cal L}_2(\alpha) = E_2(\alpha) - \mu \int |\alpha|^2
\end{equation}
which are the ${\cal O}(\epsilon^2)$ terms in the expansion of
(\ref{energy-per-particle}) and (\ref{constrained-E}), i.e. the energy
introduced in the system by the perturbation. If a diagonal Jordan form
like the one of (\ref{diagonalization}) is possible, then it is easy to
check that the second functional becomes diagonal, too
\begin{mathletters}
\begin{eqnarray}
  \label{diag-constrained-E}
  {\cal L}_2(\alpha)& = &\sum |c_k|^2 \lambda_k G(\vec{W}_k), \\
  G(\vec{W}_k) & = & \int |u_k|^2-|v_k|^2.
\end{eqnarray}
\end{mathletters}
If the stationary state, $\psi_0$, is a local minimum of the energy
subject to the constrain of a fixed norm (\ref{normalization}), then
${\cal L}_2$ must be positive for all perturbations, which has serious
implications for the eigenvalues and eigenstates. We will refer to this
later.

When studying the condensate using tools from Quantum Field Theory, one
may try a similar procedure \cite{JIST}, which is known as Bogoliubov's
theory. In that framework, the $\bar{\alpha}$ and $\alpha$ are linear
operators in a Fock space, and one searches an expansion of these
operators in terms of others that diagonalize the energy functional
(\ref{energy-perturbed}) and the evolution equations
(\ref{perturbed}). The resulting equations for the coefficients are
known as Bogoliubov's equations and correspond to the 
equations (\ref{diagonalization}) for $u_k$ and $v_k$.

\subsection{Operational procedure}

It is now useful to perform an expansion of $\alpha$ and $\bar{\alpha}$
into states of fixed vorticity so that the modes are separated into
subspaces according to their vorticities
\begin{equation}
  \vec{W}_i^{(n)} = \left( \begin{array}{c}
        u_i^{(n)}(r)e^{in\theta} \\ v_i^{(2m-n)}(r)e^{i(2m-n)\theta}
        \end{array}\right).
\end{equation}
These subspaces are not mixed by the action of the operators of
(\ref{definitions-1}), and we can define their restriction to these
subspaces
\begin{mathletters}
  \label{definitions}
  \begin{eqnarray}
    {\cal B}_n(\Omega) & = & \sigma_z{\cal H}_n(\Omega), \\
    {\cal H}_n(\Omega) & = & {\cal H}_{0n}(\Omega)+{\cal U}_n, \\
    {\cal H}_{0} & = & \left( \begin{array}{cc}
        H^{n}-(n-m)\Omega  & \\ & H^{2m-n}-(m-n)\Omega
      \end{array}\right) , \\
    H^{n} & = & -\frac{1}{2}\triangle + \frac{1}{2}(\gamma^2 r^2 + z^2)
    +\frac{n^2}{2r}+f^2-\mu(0) \\
    {\cal U}_n & = & Uf^2
    \left( \begin{array}{cc}
        1 & 1\\
        1 & 1
      \end{array}\right).
  \end{eqnarray}
\end{mathletters}

With these definitions the diagonalization procedure
(\ref{diagonalization}) becomes
\begin{equation}
  \label{eigenvalues}
  \lambda_k^n\vec{W}_k^{(n)} = {\cal B}_n(\Omega) \vec{W}_k^{(n)},
  \; n\geq m.
\end{equation}
If $G(\vec{W}_k^{(n)})>0$, then $(u_k^{(n)},v_k^{(2m-n)})$ is a
Bogoliubov mode with energy $\epsilon =\lambda_k^{(n)}$ and vorticities
$(n,2m-n)$, whereas if $G(\vec{W}_k^{(n)})<0$ then the excitation is
$(v_k^{(2m-n)},u_k^{(n)})$ with energy $\epsilon =-\lambda_k^{(n)}$. As
a rule of thumb, the $u$ function must always be the one with the
largest contribution, which is formally stated in $G(\vec{W})>0$. In the
following we will refer to these branches of the spectrum by the pairs
of quantum numbers $(n,2m-n)$ and $(2m-n,n)$, respectively.

One may find, in principle, two kinds of solutions. First, the Bogoliubov
operator may have complex eigenvalues or even have a non diagonal Jordan
form. In both cases we speak of \emph{dynamical instability} because an
arbitrarily small perturbation departures from the original state
exponentially or polynomially in time. Second, the linearized operator
may have only real eigenvalues which should be interpreted as the change
of energy in the condensate due to excitations [See Eq.
(\ref{constrained-E-perturbed})]. If $\lambda > 0$ the state under study
$\psi_0$ is a local minimum of the energy functional
(\ref{energy-functional}) with respect to this family of perturbations,
the $\lambda = 0$ case corresponds to the existence of degeneracy in the
system, and finally if $\lambda <0$ the system is told to be
energetically unstable -i.e, excitations are energetically favorable and
the state is not a local minimum of the energy.

All of the five cases exposed above have the same implications of
stability for Eq. (\ref{GPE}), which is a simple partial derivatives
equation for an order parameter, and for the more complete Bogoliubov
theory, where the perturbations are regarded as many-body corrections and
involve more degrees of freedom. Nevertheless, it must be remarked that
of the two types of instability that can be found, i.e. dynamical and
energetic instabilities, the second one is less harmful because it only
affects the dynamics when there is some kind of dissipation that drives
the system through the unstable branch. And even then the lifetime of
the system can be significant if the intensity of the destabilizing mode
is small compared to the typical times of evolution.

\subsection{Numerical procedure}

We have discretized Eq. (\ref{eigenvalues}) in a basis which is
essentially the same that we used to solve the stationary GPE. To be more
precise, the expansion is as follows
\begin{equation}
  \vec{W}_i^{(n)} = \sum _{k}a_{k}
  \left( \begin{array}{c} P_{kn}\\ 0 \end{array}\right)
  +\sum b_{l} \left( \begin{array}{c} 0\\ P_{l,2m-n} \end{array}\right).
\end{equation}
Here we have used the index convention explained above.

In this basis, the operator ${\cal H}_{0n}$ is diagonal, while the
operator ${\cal U}$ can be calculated, either by means of integrals of
the wavefunction itself in position representation, or by using a tensor
of four indices which is similar to the one in Eq.  (\ref{4-tensor}). In
any case, the equations are always linear, and so the study of the
Bogoliubov spectrum consists in building and diagonalizing a large matrix
of real numbers.

Even though the procedure is quite simple, the matrices that one must
build in order to resolve the case of strong interaction are very large
and tend to exhaust computational resources. To be able to reach a large
value of the nonlinearity we have had to work in a subspace of states
with even parity with respect to the Z axis. This way we could find the
excitations with lowest energy for different vorticities, at the cost of
missing those with odd parity, which are more energetic anyway
\cite{Note-1}.

\subsection{Analytical results}

The study explained above does not have to be performed for all of the
Bogoliubov operators in all of the possible situations. Here we will show
several important results regarding when Eq. (\ref{eigenvalues}) may
imply destabilizing modes.

{\em Lack of exponential instabilities in the Bogoliubov theory.-} Any
eigenvalue $\lambda$ satisfying Eq. (\ref{eigenvalues}) and $G(\vec{W})
\neq 0$ must be real. Eigenstates with $G(\vec{W})=0$ may involve complex
eigenvalues but they are spurious and are introduced by the linearization
procedure.

This first part is shown simply by projecting the left and right hands of
Eq.  (\ref{eigenvalues}) against the vector $\vec{W}_i^{(n)\dagger}$.
Omitting the indices the result is
\begin{eqnarray}
  \label{projection}
  \lambda_n && \int (|u|^2-|v|^2)
  = \int (\bar{u}_n H^n u + \bar{v} H^{2m-n} v) \nonumber \\
  &+& \int U f^2 |u+v|^2 - \int (n-m)\Omega (|u|^2+|v|^2).
\end{eqnarray}

The second part is more subtle. To prove it we must remember that
solutions to Eq. (\ref{perturbed}) are stationary points of the action
\cite{barrier-reson}, $S=\int L(t)dt$ corresponding to the following
Lagrangian density
\begin{equation}
  L = \int \frac{i}{2}(\alpha\bar{\alpha}_t - \bar{\alpha}\alpha_t)
  + {\cal L}_2(\alpha).
\end{equation}
Using (\ref{diag-constrained-E}) it is easy to prove that the
$G(\vec{W})=0$ modes are null modes that do not appear in the Lagrangian,
and thus are not affected by the dynamics.

It must be remarked that this result characterizes possible eigenvalues,
but does not grant that ${\cal B}_n$ have a Bogoliubov diagonalization.

{\em Sufficient condition for stability.-} If the linearized Hamiltonian
${\cal H}_n$ is positive definite, then ${\cal B}_n$ may be diagonalized,
all of its eigenvalues are positive real numbers and there are no
dynamical nor energetic instabilities.

To prove this theorem one only needs to show that there's a one-to-one
correspondence between the eigenfunctions of ${\cal
  H}_n^{1/2}\sigma_z{\cal H}_n^{1/2}$ and the eigenfunctions of
$\sigma_z{\cal H}_n$ so that
\begin{equation}
  {\cal H}_n^{1/2}\sigma_z{\cal H}_n^{1/2} |n\rangle = \lambda|n\rangle,
\end{equation}
if and only if
\begin{equation}
  \sigma_z{\cal H}_n \left({\cal H}_n^{(-1/2)}|n\rangle\right) =
  \lambda \left({\cal H}_n^{(-1/2)}|n\rangle\right).  
\end{equation}
Then one uses this result to show that the eigenvalue in
(\ref{eigenvalues}) must be positive.

{\em Stability in stationary traps.- } In Eq.  (\ref{eigenvalues}), if
$\Omega = 0$ and $n > 3m$ then the linearized Hamiltonian ${\cal H}_n$ is
positive, the Bogoliubov operator ${\cal B}_n$ can be diagonalized and it
is also positive. Furthermore, if $n > m$ then any real eigenvalue is
positive, $\lambda > 0$.

The demonstration has four steps. First, one takes any value of $n$ that
satisfies that condition and proves that $H^{2m-n} > H^m$ and $H^n > H^m
\geq 0$.  Second, this is used to prove that ${\cal H}_{0n} > {\cal
  H}_{0m}$. Third, it shown that ${\cal U}_n$ is positive which
altogether implies ${\cal H}_n>0$. The last assertion may be easily
checked with the help of (\ref{projection}).

The preceding two theorems imply that in a stationary trap any mode with
negative energy must be comprised in the $(0,-2m),\ldots,(m,m)$ families,
and any dynamic instability must lay in $(0,-2m),\ldots,(3m,0)$. Thus we
need only diagonalize a finite number of operators to make sure the
system is stable or unstable. This result is an extension of the one
obtained in Ref. \cite{Fetter}, where a sufficient condition for
stability is found to be $n^2 \geq 4m^2$, without taking into account
possible complex eigenvalues.

{\em Local stability under rotation.-} In Eq. (\ref{eigenvalues}) the
operator ${\cal B}_n(\Omega)$ exhibits a linear dependence with respect
to $\Omega$
\begin{equation}
  \label{bogol-shift}
  {\cal B}_n(\Omega) = {\cal B}_n(0) - (n-m)\Omega.
\end{equation}
While the wave functions of the modes are the same as the ones of the
stationary traps, the energies of the excitations suffer a global shift
that depends on the vorticity
\begin{equation}
  \lambda(\Omega) = \lambda(0) - (n-m)\Omega.
\end{equation}

In general, the influence of these shifts has to be checked numerically.
It is easy to show, however, that the shift is positive for $n < m$,
which means that the possibly negative eigenvalues in the range $0 < n <
m$ can be suppressed if $\Omega$ is large enough. Even more, as the shift
is a real number, if one demonstrates that there are no dynamical
instabilities in the stationary trap, then there will be no dynamical
instabilities in the rotating trap, neither.

\subsection{Numerical results}

Summing up, from a practical point of view, the issue of stability
consists in two different steps. The first one is the search for a
stationary solution of the GPE with the appropriate vorticity, which we
have already performed in Sect. \ref{sec:stationary}, and the second one
is the study of the spectrum of the Bogoliubov operators for this
particular state.

{\em Stability of the $m=1$ vortex-line in a stationary trap.-} In this
case of unit charge one only has to study a single operator, ${\cal
  B}_0$, to know whether the system is stable. This calculation provides
us with the branch of the spectrum of excitations which is characterized
by the quantum numbers $(0,2)$ and $(2,0)$, as already explained. We have
done this for a wide range of nonlinearities in the absence of rotation,
$\Omega =0$, and the first conclusion is that the Bogoliubov operator has
a diagonal Jordan form with all eigenvalues being real.

\begin{figure} {\centering
    \resizebox*{0.9\columnwidth}{!}{\includegraphics{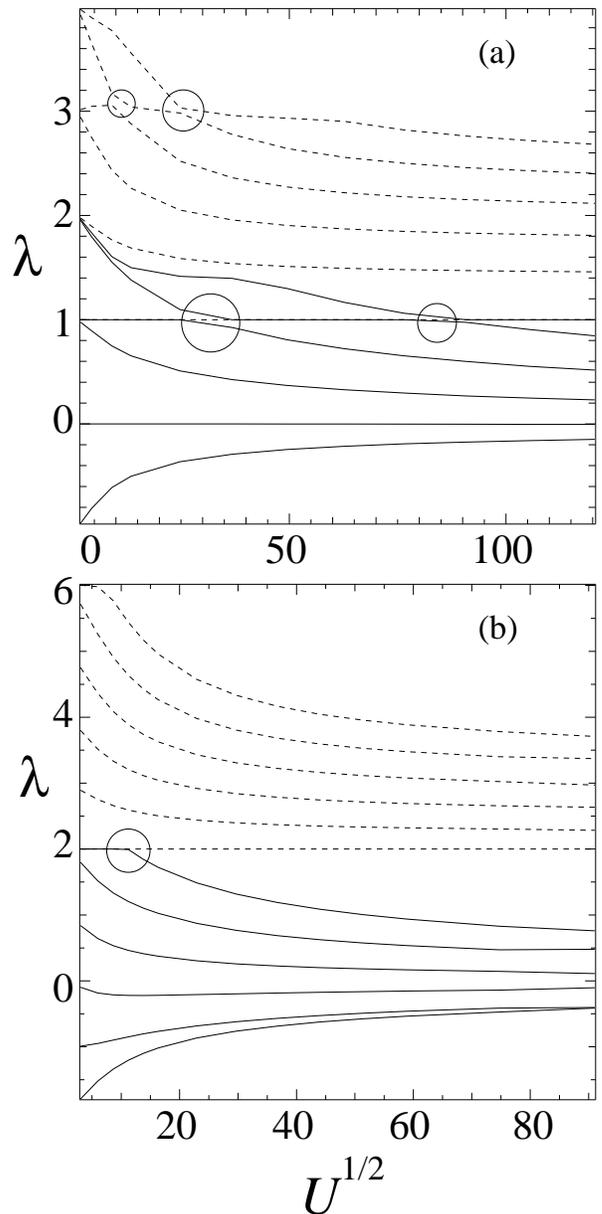}}
    \par}
  \caption{\label{spec}
    Lowest eigenvalues of the Bogoliubov operator \protect${\cal
      B}_0\protect$ for the \protect$m=1\protect$ unperturbed state in
    (a) a spherically symmetric trap, \protect$\gamma =1\protect$, and
    (b) an axially symmetric trap, \protect$\gamma =2 \protect$.  The
    solid lines represent modes with quantum numbers
    \protect$(0,2)\protect$ and the dashed lines represent modes of the
    \protect$(2,0)\protect$ family.  Crossing of levels is signaled with
    circles as a visual aid.}
\end{figure}

\begin{figure} {\centering
    \resizebox*{0.9\columnwidth}{!}{\includegraphics{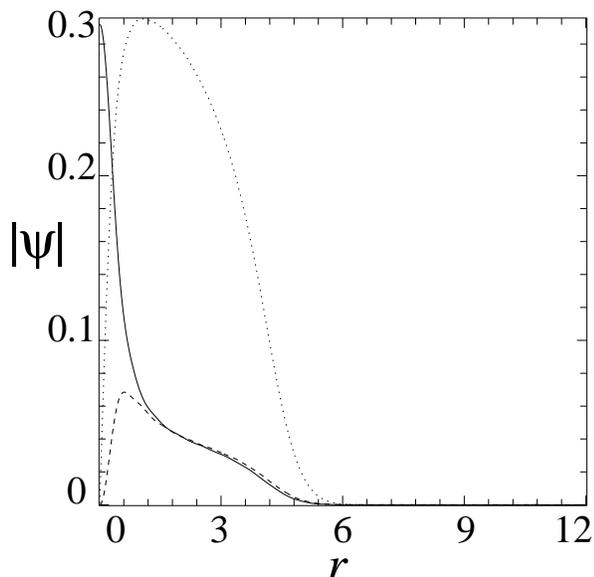}} \par}
  \caption{\label{Waves} Shape of the destabilizing mode. We show both
    the original solution (dotted line), the largest contribution,
    \protect$ u\protect $, (dashed line) and the smallest
    contribution, \protect$ v\protect $, (solid line). Functions
    have been rescaled to aid visualization.}
\end{figure}

In Fig \ref{spec} we show a selected set of the eigenvalues of the
Bogoliubov operator, both for a spherically symmetric trap and an
elongated trap. In those pictures one sees several things.  First, there
are two constant eigenvalues $\lambda =1$ which correspond to
oscillations of the vortex line along the Z axis.  Second, there is a
single neutral mode $\lambda =0$ only for the spherically symmetric trap,
which corresponds to the symmetry of rotation of the condensate around an
axis on the XY plane. The symmetry and the mode disappear when $\gamma =
2$ [See Fig. \ref{spec}]. And finally there is at least one negative
eigenvalue $\mu <0$ (more in the case of an elongated trap) which is
responsible for the energetic destabilization of the system. The largest
contribution to this destabilizing mode is a wavefunction captured in the
vortex line and has zero vorticity (i.e. it is a \emph{core} mode) [See
Fig.  \ref{Waves}] as it was qualitatively predicted by Rokhsar in Ref.
\cite{Rokhsar}.

We must remark that the number of unstable modes increases with the
geometry factor: the more elongated the trap is, the easier is to
transfer energy from the vortex to the core plus longitudinal
excitations. In other words, for $\gamma \leq 1$ (spherical or
``pancake'' traps) there's only one negative eigenvalue which corresponds
to an excitation with a different vorticity than the unperturbed
function, while for $\gamma \geq 1$ we still have that mode, plus some
more which are excited with respect to the Z axis. As a consequence, if
the experiment is subject to dissipation and these unstable modes play a
significant role in the dynamics, then the more elongated the trap is the
less stable the vortex will be.

\begin{figure} {\centering
    \resizebox*{0.9\columnwidth}{!}{\includegraphics{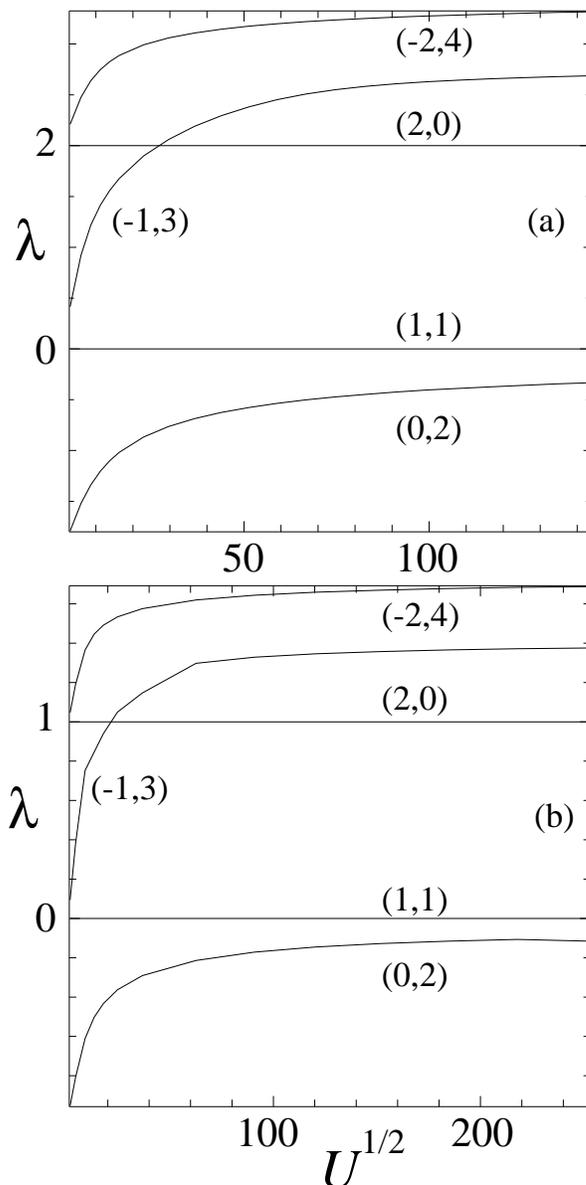}} \par}
  \caption{\label{split}Excitation energy of the lowest Bogoliubov
    modes for an unperturbed state with \protect$m=1\protect $. The
    vorticities of each mode is written close to its corresponding
    line. Plot (a) corresponds to $\gamma = 1$, and (b) to \protect$
    \gamma = 2\protect $.}
\end{figure}

In Fig \ref{split} we also show the lowest eigenvalues of the families
$(-1,3)$,$(1,1)$, $(0,2)$, $(2,0)$ and $(-2,-4)$, that is, excitations
where the main contribution is an eigenstate of $L_{z}$ with eigenvalues
$m=0,\pm 1,\pm 2$.  In those pictures one sees that subspaces with
excitations of the same vorticity but opposite sign have also different
energy, a phenomenon which is solely due to the interaction.

{\em Stability of $m=1$ vortex lines in rotating traps.-} It was already
proved (\ref{bogol-shift}) that, as the effect of rotation is
gradually turned on, the modes with $n<m$ and with $n>m$ are shifted up
and down in the spectrum, respectively. It remained the question of
whether the shift is enough to stabilize the vortex states, and the
answer is yes, according to numerical experiments.

\begin{figure} {\centering
    \resizebox*{0.9\columnwidth}{!}{\includegraphics{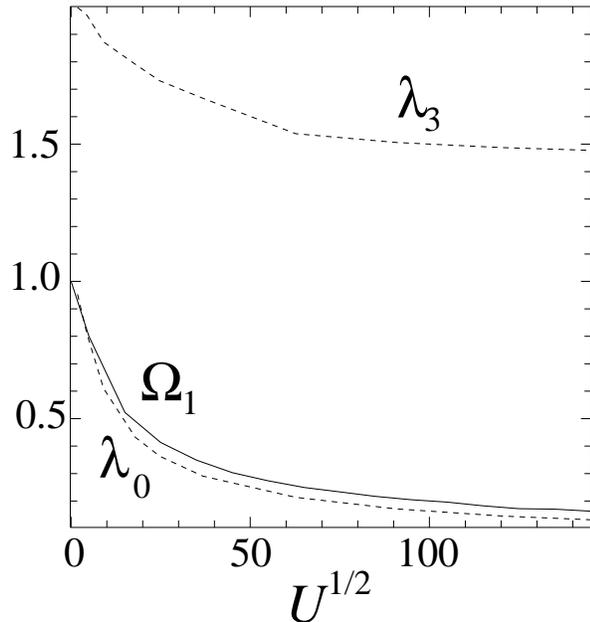}} \par}
  \caption{\label{stab-rot}
    The solid line, the lower dashed line and the upper dashed line are
    $\Omega_1$, $-\lambda_0$ and $\lambda_3$, respectively, for a
    spherically symmetric trap, $\gamma=1$.}
\end{figure}

First, as it is shown in Fig. \ref{stab-rot}, the negative eigenvalue is
slightly smaller than the stabilizing frequency, $|\lambda_0| <
\Omega_1$, which implies that for $\Omega > \Omega_1$ the energetically
unstable branch with vorticity $m=0$ disappears. And second, the
eigenvalues of ${\cal B}_n$ for $n > m$ are found to be larger than
$(n-m)\Omega_1$. In consequence for at least the interval
$[\Omega_1,\Omega_2)$ all of the ${\cal B}_n$ operators are positive and
the vortex with unit charge is a local minimum of the energy functional.

In any case the shifts are always real, which implies that the ${\cal
  B}_n$ operators remain diagonalizable with real eigenvalues and without 
dynamical instabilities.

{\em Stability of the $m=2$ vortex line.-} Another interesting
configuration is the $m=2$ multicharged vortex line. Here one suspects
that a configuration with several vortices of unit charge has less energy
than a single multicharged vortex, under all circumstances. In other
words, they must be always energetically unstable.

This intuitive perception is confirmed by the numerics. First the
diagonalization of ${\cal B}_1$ reveals that this operator has at least
one negative eigenvalue, while ${\cal B}_0$ has both negative eigenvalues
and a pair of complex eigenvalues that, as we saw above, do not
participate in the dynamics and must be discarded. Regarding the negative
eigenvalues, they do not decrease with the nonlinearity, but are always
larger in absolute value than their linear limits. This implies that
there are always negative eigenvalues which cannot be suppressed with any
rotation below the critical value, $\Omega_c > \Omega > \Omega_2$.

The immediate consequence of this linear stability analysis is that, due
to the linearization of the energy (\ref{energy-perturbed}) not being
positive, the $m=2$ vortex-line is never a local minimum of the
energy. This is true even for the parameter interval,
$[\Omega_2,\Omega_3)$, in which it has less energy than the rest of
stationary states of well defined symmetry. If the $m=2$ ground state is
not a minimum, and the other symmetric states have more energy, we can
conclude that the minimum of the energy functional in the rotating trap
with $\Omega \in [\Omega_2,\Omega_3)$ must be a state which is not
symmetric with respect to rotations \cite{Rokhsar2}. A similar analysis
can be performed for the stationary states with $m=3,4\ldots$ which
extends this result to larger rotation frequencies, all below the
critical one.

\subsection{Lyapunov stability}

Speaking roughly, a solution of Eq. (\ref{GPE}) is Lyapunov stable when
every perturbed solution which is close enough to the original wave
remains close throughout the evolution. The concepts of Lyapunov
stability and linear stability are close, but the latter does not imply
the former as it is only defined in the limit of infinitesimal
perturbations.

Studying the Lyapunov stability of Eq. (\ref{GPE}) theoretically is a
difficult task that should be subject of further investigation. In the
mean time we have performed an ``empirical'' study of the Lyapunov
stability of the stationary solutions with $m=1$ and $m=2$ vorticities,
by simulating numerically how they evolve for small perturbations and
long times. The simulation was performed with a three-dimensional
split-step pseudospectral method like the one from
Ref. \cite{barrier-reson}, using a $80 \times 80
\times 80$ points grid to study both the $\gamma=1$ and $\gamma=2$
problems.

The main result of this complementary work is that both the unit charge
vortex line and the multicharged vortex line are stable to perturbations
which involve the destabilizing modes as defined by (\ref{eigenvalues}).
For example, one may try to add a small contribution ($0.5\%$) of a core
mode to the $m=2$ vortex, and with the result that the vortex line is
split into two unit charge vortex lines, which rotate but remain close to
the origin. We must remark that, although these simulations only work for
finite times which are dictated by the precision of the scheme and the
computational resources, these times are typically 20 or 30 periods of
the trap, which is much larger than any of the magnitudes that one may
address theoretically to the destabilization process (i.e. the negative
or complex eigenvalues of Eq. (\ref{eigenvalues}))

In the end, what this type of simulations reveal is that the $m=1$ and
$m=2$ stationary states are energetically unstable, but this has no
influence on the dynamic unless some other ``mixing'' or dissipative
terms participate in the model.

\section{Conclusions}
\label{sec:conclusions}

We have studied the vortex solutions of a dilute, nonuniform Bose
condensed gas as modeled by the Gross-Pitaevskii equation (\ref{GPE}),
both in a stationary, axially symmetric trap, and subject to rotation (or
a uniform magnetic field).

First, we have searched solutions of Eq. (\ref{stat-eq}) that have the
lowest energy and which are also eigenstates of the third component of
the angular momentum operator, $\psi (r,z,\theta )=f(r,z)e^{im\theta}$,
both in a stationary trap and in a rotating trap, and from small to very
large nonlinearities. It has been found that a nonzero angular speed (or
magnetic field) is necessary in order to turn a vortex line state into a
minimum of the energy functional with respect to other states of
well-defined vorticity. However it remains open the question of whether
the minimum of energy must have a well defined vorticity.

Next we have studied the stability of these stationary solutions of the
GPE.  We have formulated a set of coupled equations that describe both
the linearization of the GPE around a stationary solution and
Bogoliubov's corrections to the mean field theory that describes the
condensate. It has been proved that the problem may not exhibit dynamical
instabilities of exponential nature, plus several other theorems that
describe the phenomenology associated to the possible instabilities.

The perturbative equations have been solved numerically for stationary
states having $m=1$ and $m=2$ vorticities. In both cases it has been
found that the only instability is of energetic nature, being limited to
a small number of modes whose nature had already been predicted in
\cite{Rokhsar}.

For the vortex with unit charge we have found that this instability may
be suppressed by rotating the trap at a suitable speed, and even when the
trap is stationary, it is expected that it plays no significant role in
the dynamics unless there is enough dissipation as to take the system
through the unstable branch. On the other hand, the linear stability
analysis for the $m=2$ multicharged vortex reveals that the energetic
instability may never be suppressed, and that this configuration is never
a minimum of the energy functional, even though its lifetime is, once
more, only conditioned by possible dissipation.

The last and probably most important conclusion of this work is that in
the rotating trap, and for $\Omega > \Omega_2$, the state of minimum
energy is not an eigenstate of the $L_z$ operator, and thus it is not
symmetric with respect to rotations. A similar result has been found in
Ref. \cite{Rokhsar2} by means of a minimization procedure which is only
justified in the limit of very small $U$, while our demonstration remains
valid for all nonlinearities, as far as the linearization procedure may
be carried on.

From a experimental point of view, this work has several implications.
First, it is clear the conclusion that vortex lines with unit charge may
be produced by rotating the trap at a suitable speed and then cooling the
gas. Second, once rotation is removed, these vortices will survive for a
long time if dissipation is small. Third, the multi-charged vortices are
not minimum of the energy functional and thus it will be difficult to
produce them by mean of cooling a rotating gas. And finally, if these
multi-charged vortices are produced by some other mean such as Quantum
Engineering, then we can assure that their lifetime will only depend on
the intensity of dissipation, whose effect is to take the system either
to the $m=0$ ground state if $\Omega<\Omega_1$, to the unit charge
vortex-line state if $\Omega < \Omega_2$, or to a symmetry-less
multicharged state if $\Omega > \Omega_2$ (A phenomenon which is regarded
as splitting in the literature).

The numerical results found in the paper have been possible due to the
use of a powerful Galerkin spectral method optimized to allow the
consideration of thousands of modes which is a step forward with respect
to the previous analysis.

\acknowledgements

This work has been partially supported by DGICyT under grants PB96-0534
and PB95-0389 We thank Prof. Cirac from the Institud f\"ur Theoretische
Physik of Innsbruck for proposing us the problem and helping in the
development of the stability theorems.


\begin{thebibliography}{}
\bibitem{Precision}{This kind of solutions are sometimes called
    eigenstates or solitary waves; we prefer the term stationary state
    since the term eigenstate is more properly used in the framework
    of linear equations.}
\bibitem{Note-1}{This has to do with the fact that states of even parity
    have less zeros than states with odd parity.}
\end{thebibliography}
\end{document}